# University Rents Enabling Corporate Innovation: Mapping Academic Researcher Coding and Discursive Labour in the R Language Ecosystem

XIAOLAN CAI[1]

MATHIEU O'NEIL

News and Media Research Centre, University of Canberra, Australia

STEFANO ZACCHIROLI

LTCI, Télécom Paris, Institut Polytechnique de Paris, Palaiseau, France

This article explores the role of unrecognised labour in corporate innovation systems via an analysis of researcher coding and discursive contributions to R, one of the largest statistical software ecosystems. Studies of online platforms typically focus on how platform affordances constrain participants' actions, and profit from their labour. We innovate by connecting the labour performed inside digital platforms to the professional employment of participants. Our case study analyses 8,924 R package repositories on GitHub, examining commits and communications. Our quantitative findings show that researchers, alongside non-affiliated contributors, are the most frequent owners of R package repositories and their most active contributors. Researchers are more likely to hold official roles compared to the average, and to engage in collaborative problem-solving and support work during package development. This means there is, underneath the 'recognised' category of star researchers who transition between academia and industry and secure generous funding, an 'unrecognised' category of researchers who not only create and maintain key statistical infrastructure, but also provide support to industry employees, for no remuneration. Our qualitative findings show how this unrecognised labour affects practitioners. Finally, our analysis of the ideology and practice of free, libre and open source software (FLOSS) shows how this ideology and practice legitimate the use of 'university rents' by Big Tech. In conclusion, we argue that existing mechanisms are insufficient to ensure these digital commons' sustainability: FLOSS needs broader systemic support.

[1] Corresponding author : Xiaolan Cai <Xiaolan.Cai@canberra.edu.au>
Date submitted: 30 September 2025

## Introduction

Increasing scholarly attention is being paid to the role of researchers in the development of 'corporate innovation systems' (Rikap, 2024). There is a growing flow of researchers from academia to industry, particularly from elite institutions into technology companies such as Google, Microsoft and Meta. A reconstruction of the affiliation history of over 60,000 computer science researchers found that 8% had moved from academia to industry, with a sharp increase in transitions in the last decade, particularly for those working on machine learning with higher citation rates (Jurowetski et al., 2021).

Data scientists and engineers working for Big Tech firms can be seen as a bridge between software co-production and appropriation, because Big Tech firms not only develop software in-house but also work with universities, public research organisations and other firms in joint research projects. These collaborations result in thousands of co-authored papers but only a dozen co-owned patents, because most of the thousands of Big Tech patents are exclusively owned. In the case of Microsoft for example, 78,3% of its 17,405 publications between 2014 and 2019 were co-authored with university researchers; during the same period Microsoft applied and was granted 76,109 patents, 0,2% of which were co-owned (Rikap & Lundvall, 2022). This appropriation has been described as a 'predatory' (Rikap & Lundvall, 2022) type of 'intellectual monopoly' (Pagano, 2014).

Less attention has been paid to the role of researchers in creating and maintaining critical digital infrastructure, such as statistical software. In this article, we explore the role of researcher labour in the contemporary platform economy via an analysis of researchers' coding and discursive contributions to R, one of the largest statistical software ecosystems. Data science and statistical software are not just used by researchers: before any data-based machine learning can occur, these data need to be ordered, curated, and analysed.

We first provide an empirical mapping of the software ecosystem of free, libre and open source software (FLOSS) packages built around R on the GitHub platform, which enables diverse participants to work collaboratively. Using descriptive statistics of commits to R package repositories, we focus on the role of academic researchers, and our first research question is:

> RQ1: How much do academic researchers own (create or inherit from others) R package repositories on GitHub, in relation to other categories of contributors, e.g., firm employees, foundation employees, etc.?

We next focus on a key part of collaborative software development, *committing* (i.e., contributing new code or modifying existing code by submitting an update to a version control system repository), and our second research question is:

> RQ2: How much do researchers commit to R package repositories on GitHub, in relation to other categories of contributors, e.g., firm employees, foundation employees, etc.?

To answer, we draw on metrics developed by O'Neil et al. (2024) to map contributions to GitHub. We augment these methods by conducting a quantitative content analysis of *issues*, communication channels on GitHub which are used to plan, discuss and track work in repositories, thus investigating not just 'who does what,' but also '*who asks who* to do what'? This enables us to explore the communicative dimension of open source maintaining, e.g., managing users by providing feedback about their contributions as well as (whenever needed) support. We analyse the roles of different categories of participants in creating issues, assigning issues to others, and discussing issues, so our third research question is:

> RQ3: What kind of issues do researchers engage with, in relation to other categories of contributors, e.g., firm employees, foundation employees, etc.?

Our analyses provide new insights into the relationship between academia and industry, and volunteer and paid labour. Specifically, our research advances the understanding of the political economy of 'intellectual monopoly capitalism' (Pagano, 2014) in three ways: first, our quantitative findings show that there is, underneath the 'recognised' category of star computer science researchers who transition between academia and industry and secure generous funding, an 'unrecognised' category of researchers who not only create and maintain key statistical infrastructure, but also provide support to industry employees, for no remuneration; second, our qualitative findings show how this labour affects practitioners; third, our analysis of the ideology and practice of free, libre and open source software (FLOSS) shows how this ideology and practice legitimate the unrecognised use of 'university rents' by Big Tech firms.

The article is structured as follows. After summarising the literature on Big Tech and research, as well as FLOSS and industry, we present our case study, methods, and findings. In our discussion and concluding remarks, we outline how Big Tech's extraction of university rents is legitimated by intrinsic aspects of FLOSS ideals and outline corrective policies.

## Literature review

### *Big Tech and research*

The rise of Big Tech-funded research is part of a broader trend, the decline in public investment in research. Between 1981 and 2013 the share of public-financed R&D to GDP in OECD countries was reduced from 0.82% to 0.67%. By contrast, industry-financed R&D increased from 0.96% of GDP in 1981 to 1.44% in 2013 (Archibugi & Filippetti, 2018). The traditional division between publicly-funded 'fundamental' and privately-funded 'applied' research no longer applies (Jurowetzki et al. 2021). Analysts focus on the strategies employed by Big Tech firms to enrol researchers (Rikap, 2024), which facilitates their ecosystem lock-in; this creates quasi-insurmountable obstacles to competition and enhances firms' profits and role in the economy (Jacobides et al., 2021).

There are two risks associated to the co-optation of research by the private sector. In terms of production, there is no guarantee that market-led opportunities correspond to societal needs and priorities. Innovations developed outside any consideration for the public interest will favour short term profit over social usefulness (Jurowetzki et al., 2021). Of particular concern is the increasing tendency to leverage digital footprints, so that personal data become part of gigantic datasets, which are then analysed to refine Big Tech algorithms and keep potential competitors at bay (Rikap, 2021). In terms of dissemination, the excessive privatisation of knowledge reduces the possibilities of diffusion. Indeed public and private-generated knowledge have different economic characteristics, particularly in terms of their degrees of rivalry and excludability (Archibugi & Filippetti, 2018).

### *The adoption of free, libre and open source software by industry*

Internet infrastructures rely heavily on free, libre and open source software (FLOSS) and associated projects and communities. To understand why FLOSS has been adopted by industry, we must remember that the design of such non-rival digital commons is directed by the requirements of the communities which produce them; the objectives and outcomes of a project primarily reflect the demands of initial contributors, who are also its prime beneficiaries. This integrity of product and process results in exceptional robustness.

Firms negated the disruptive potential of FLOSS and ensured continuous participation by acceding to its proponents' principal demand: providing access to code via open licenses. This was an evolution between what Sebastien Broca (2018) defined as Microsoft's 'informational capitalism' – the firm seeks to protect the value of its closed proprietary systems – to Google's more flexible 'digital capitalism' – the firm integrates the commons into its business model and prioritises mobile and cloud business models using big data and artificial intelligence.

The FLOSS collaborative development model was first embraced by Red Hat in 1993 and IBM in 2002. Google's Android, launched in 2008, is based on Linux, and Microsoft bought the GitHub development platform for $7.5 billion in 2018; FLOSS constitutes up to 90% of software stacks.[2] Researchers have estimated that without FLOSS to power digital products and services, it would cost companies $8.8 trillion to develop this software from the ground up (Hoffmann et al., 2024). Engaging with FLOSS is now also a source of legitimacy and enables firms, in the highly competitive skilled IT professional job market, to attract prospective employees; thanks to coordination by foundations – chief amongst which is the Linux Foundation (Biddle, 2019) – firms can share development costs with other firms; finally firm investment may also lead to shaping the governance and technical orientation of software projects and products (Muselli et al., 2024).

---

[2][1] See Synopsys, 2020 Open Source Security and Risk Analysis Report" https://www.synopsys.com/content/dam/synopsys/sig-assets/reports/2020-ossra-report.pdf; Sonatype 2020 State of the Software Supply Chain. https://www.sonatype.com/resources/white-paper-state-of-the-software-supply-chain-2020

### Analyses of GitHub as a 'social coding' platform

A key site in the industry-FLOSS relationship is GitHub, a hosting platform based on the Git version control system which makes participation in FLOSS projects highly visible. GitHub allows users to host free copies of software source code in software 'repositories' and, provided that the owner of the receiving repository allows it, to exchange code improvements among related repositories. Repositories typically comprise a coding project's files, revision history and participant discussion history. GitHub has grown to become the most popular FLOSS collaborative development platform, hosting more than 339 million public repositories,[3] and becoming the 'cultural epicentre' for open source's rapid growth (Eghbal, 2016). GitHub is a site of collaborative production and a key coordination hub between firms and projects in the global platform ecosystem; until recently, it was not primarily used by firms to extract and monetise user data. However in 2018 Microsoft purchased GitHub for $7.5 billion and now trains its GitHub Copilot generative AI software, which is a paid service, on the content of GitHub's public repositories. In addition, the way users populate their personal pages with their accomplishments and 'follow' one another are not dissimilar to that of social media platforms. Measures of popularity include users awarding 'stars' to repositories to indicate their appreciation for their quality, and account holders electing to be notified when changes are made to repositories, becoming known as 'watchers'. According to one study, users attribute increased participation on their projects to GitHub features: with 'pull requests,' people participate in and observe the complete unfolding of the collaborative development process, providing transparency and feedback to submitters, which enhances continuing contributions (McDonald and Goggin, 2013). Similarly, hoc analyses performed by community members enable them to make informed assumptions such as deducing someone's technical goals and vision when they edit code, or which of several similar projects has the best chance of thriving in the long term (Dabbish et al., 2013). A survey of 791 GitHub developers found that 73% considered the number of stars before contributing to or using a project (Borges and Valente, 2018).

---

[3] See https://innovationgraph.github.com/global-metrics/repositories

Studies focusing on these internal platform mechanisms do not address how activity on GitHub relates to contributors' employment outside the platform; nor do they explore from where value is extracted, and for whose benefit. To date political economy investigations of social coding have focused on two issues: labour activism and firm involvement. In the first case, GitHub's key industrial function means it faces less censorship in authoritarian countries than other platforms. It was used by mainland Chinese developers to discuss overwork in the '996-ICU' repository, so named in reference to the fact that being expected to work from 9am to 9pm six days a week would 'lead developers to the Intensive Care Unit' (Zhen, 2021). In the second case, researchers have sought to account for the influence of firms such as Microsoft and Google in GitHub's most active repositories. These firms have become dominant contributors, via employees who are paid to develop open source code. Firm employee work on GitHub was investigated include using email domain names as proxies for employment. This led to the discovery of 'contribution territories' whereby significant contributions of employees of dominant firms never co-occurred in key repositories (O'Neil et al., 2024).

# CASE STUDY, METHODS AND DATASETS

## *Contributions and communications on GitHub*

A *repository* is a place in GitHub to store code, files, and each file's revision history. GitHub repositories comprise different official *roles* which are assigned to developers to reflect ownership, decision-making power and repository management responsibilities. From core to peripheral, roles include *owners* who have full control over the repository; *members* of the organisation which owns the repository have enhanced levels access to a repository; *maintainers* oversee the health of the repository, review contributions and manage issues; *collaborators* are external participants who are invited to contribute; and *contributors* submit code, issues, or discussions but do not have direct repository management responsibilities.

A *commit* records changes to one or more files in a branch (a snapshot of a repository). A *commit author* is the person who originally wrote the code or made the change, whilst a *committer* is the person who last applied the code or change to the repository; in collaborative projects, the commit author and committer could be different.

Communication is an essential aspect of self-directed, non-hierarchical, collaborative production (Dafermos, 2021). Communication on GitHub primarily occurs via *issues*. Issues enable participants to plan, discuss, and track work regarding repository details. Issue *comments* are features enabling users to discuss issues. Participants report new issues by providing a *title* and an optional *description* of the issue. As issues of different types (e.g., asking questions, proposing features, reporting bugs) and quality can be submitted, GitHub offers a customisable *labelling* system that can be used to tag issue reports, enabling GitHub developers to categorise issues, pull requests and related development tasks; such labelling has positive effects on issue processing (Liao et al., 2018). GitHub provides nine default labels (e.g., 'bug,' 'enhancement' and 'documentation') and allows developers to create their own labels; more than one label can be attached to an issue. Finally, an issue *assignee* is the person an issue is assigned to; they are responsible for progressing and fixing it.

*Site selection*

We focus on R, a programming language mainly used for data analysis and developing statistical software released under the GNU General Public License (GPL). The core R language is augmented by many extension packages containing reusable code and documentation. R is popular with researchers, but is also used by firms for statistical analysis, data analysis and visualisation, and machine learning. Tracking precise industry use of FLOSS is challenging (Eghbal, 2016). In the case of R, a clear sign of industry interest and use can be inferred from the list of R Consortium sponsors.[4] These include pharmaceutical companies (e.g., Pfizer), biotechnology companies (e.g., Roche), technology companies (e.g., Microsoft, Google and Esri), data analytics consultancies (e.g., Lander Analytics) and insurance companies (e.g., Swiss Re). A complete list is available in Appendix 1.

The Comprehensive R Archive Network (CRAN), launched in 1997 is the official distribution platform of R packages with more than 100 mirrors worldwide.[5] In May of 2022 there were 18,600 R packages distributed via CRAN. A decade after the launch of CRAN, Bioconductor was established to distribute bioinformation R packages (2,100 packages in May 2022). R-Forge, a centralised web-based R package development platform, was established in 2007 (2,138 packages registered in May 2022). While the number of packages registered on R-Forge has increased consistently since its establishment, GitHub has become a popular development host in the last decade both among packages distributed on CRAN and Bioconductor. To understand the contributions and roles of academic researchers in R packages development, we focused on CRAN packages which had repositories on GitHub, including mirrors from other development platforms such as R-Forge.

---

[4] The R Consortium seeks to 'advance the worldwide promotion of and support for the R open source language.' See https://r-consortium.org/about/

[5] Mirrors are copies of the entire set of R packages created to increase data availability and resilience.

*Data collection*

We used the "tools::CRAN_package_db" function in R in May 2022 to generate a list of R packages (18.6K) on CRAN. To capture as broad a picture as possible we used the GitHub REST API to generate another list of R packages (22.3K) from 'https://github.com/cran' (an unofficial mirror of CRAN on GitHub containing archived packages which may no longer be distributed via CRAN). We then combined the two lists by removing duplicates, through identification of unique package names, obtaining 23,270 packages in total.

Next, we investigated the 'URL,' 'BugReports' and 'about' fields of these packages to identify their online repositories' URLs. We found that 39% (9,086) of these 23,270 packages were hosted in 8,924 GitHub repositories, 12% (2,701) are hosted on other platforms and 49% (11,471) did not provide URLs of online repositories in their profiles (Table 1). While most of the 8,924 GitHub repositories only hosted 1 package, 86 repositories hosted more than one packages (ranging from 2 to 14, Table 2).

**Table 1. Summary of R packages on CRAN, unofficial CRAN mirror on GitHub, and their online repositories' URLs**

| **Source** | **Has GitHub repository URL** | **Doesn't have GitHub repository URL** | | **Total** |
| --- | --- | --- | --- | --- |
| | | Has URL of repository on other platforms | Has no URL of on-line repository | |
| **Both** | 8,023 | 2,220 | 8,289 | 18,544 |
| **Only CRAN** | 35 | 1 | 34 | 70 |
| **Only Mirror** | 1,028 | 480 | 3148 | 4,656 |
| **Total** | **9,086** | **2,701** | **11,471** | **23,270** |
| | **39%** | **12%** | **49%** | **100%** |

**Table 2. Distribution of number of packages hosted by a repository**

| Number of R packages in a repository | Count |
|---:|---:|
| 1 | 8838 |
| 2 | 56 |
| 3 | 16 |
| 4 | 6 |
| 5 | 3 |
| 6 | 1 |
| 8 | 1 |
| 9 | 1 |
| 12 | 1 |
| 14 | 1 |
| **TOTAL** | **8,924** |

Using our list of 8,924 repositories, we collected a dataset of commits and a dataset of issues.

*Dataset 1: Commits*

We used the Python package 'GitPython' (Trier, 2016) to clone R-focused repositories, and extracted all commit information (N:2,339,669) from these repositories with 'PyDriller' (Spadini et al., 2018) via Google Colab. There is no rate limit to collect commits information, so we obtained information for all the commits of the 8,924 repositories. Git is an open source version control system software, independent from specific platforms. GitHub is a platform offering Git repository hosting services (gratis or not). GitHub usernames are GitHub-specific, only known to GitHub and not recorded in Git repositories; in turn Git only records usernames and email addresses, as configured by Git users on their systems.

The collected commits dataset includes commit ID (SHA), commit message (a brief description of the code or the change); commit author's name, email address, date and time they created the code or made the change; committer's name, email address, date and time the code or change was added to the repository; and source line of code (SLOC, total number of added and deleted lines in the commit). There were 23,339 code contributors (6,589 as 'author' only, 70 as 'committer' only, and 16,680 as both 'author' and 'committer') who participated in the development of these 8,924 repositories.

*Dataset 2: Issues*

We collected 285,395 issues from 6,746 repositories via the GitHub REST API using the 'rvest' R package (Wickham, 2016). In contrast to commit harvesting, the GitHub REST API limits the collection of issues to 5,000 per hour; we used the 'Sys.sleep(3600)' function to 'throttle' or suspend data collection, in this case for an hour, when the maximum number was reached, before resuming data collection. This enabled us to collect all the issues in the 6,746 repositories. We found that 44,948 unique users created issues. There were 2,178 repositories without any issues. We found that 75% of the issue-less repositories had no more than three stars, no more than two watchers, and no more than twelve forks, suggesting that those repositories had relatively low popularity and that issues form an important part of the development of successful repositories.

The collected issue dataset included the ID of issues, the repository an issue was related to, the issue's title, labels associated to the issue, the date an issue was created and (if applicable) closed, the issue creator's username, the number of comments posted to an issue, the URL of these comments, the issue assignee's username, and the user who closed the issue's username.

In addition to issues, we also collected 613,793 issue comments (e.g., 'please provide additional information'). We collected the usernames of commenters, the time and date they commented and the content of their comments. There were 98,663 (34.57%) issues without comments; for the issues with comments, the mean number of comments was 3.3. We detected the languages used in issues' title and text as well as comments text by using the 'cld2' R package, and found that most of the issues (N:246,338, 86.31%) and issues' comments (N:585,909, 95.46%) were in English. In our data analysis, we only focus on issues and comments written in English.

### *Data processing: user affiliation identification and classification*

To understand the distribution of contributions and communications between different professional categories of participants, we needed to identify their affiliations. We followed O'Neil et al. (2024) by using email address domains such as 'uni.edu.au' and 'firm.com' derived from email addresses as a proxy for user affiliation, where '@fb.com' is a proxy for Facebook.

Since the names in our commits dataset are different from GitHub usernames (cf. 'Dataset 1: commits' section above), and email addresses could not be collected in our issue dataset via the GitHub REST API, we did not link these two datasets by usernames nor email addresses in our analysis. Instead, for the commits dataset, we used email address domains as a proxy for user affiliation, whilst we used both affiliations from user pages and email address domains to determine affiliations of users in the issues dataset. The same person can author commits in our dataset under different digital identities, in particular different email addresses. Given our study focuses on contribution categories, we did not apply traditional techniques of identify merging (Goeminne and Mens, 2013). This allows to account separately for commits made using a professional email address (corporate, academic, etc.) and using a personal one.

We then obtained a list of university domains from Wikipedia and WikiData data. We also obtained users' profiles including their affiliations, email addresses, and location by using all the usernames of contributors in issues (N:36,411) with the GitHub REST API using the 'rvest' R package. We linked the affiliations provided by users in their profiles (when available) with email addresses (when available) to inform our list of employers and obtained a list of 1,091 pairs of unique email domains and affiliations (including email domains of email servers such as 'gmail.com' and personal domains such as 'JohnDoe.me') and classified them into five categories: research institutions, firms, governments, foundations, and non-affiliated. Since they did not give credit to their employers by using organisational email addresses, we classified contributors who used their personal website email address instead of a corporate (private or public) email address as 'non-affiliated' even if we could determine some of these contributors' affiliations through their personal websites. Then we merged the list of universities, and their domains (derived from Wikipedia and WikiData data), and the list of affiliations and their email address domains from users' profile.

We used this 'merged list' (university domains and affiliations and email addresses from user profiles) to connect email domains in our commits dataset. In the case of email address domains that did not exist in the 'merged list,' we manually checked them using Web searches and added these new entries to this augmented 'merged list.' Finally, we classified the remaining 306 email address domains whose affiliation could not be identified as 'non-affiliated' (N:306, 4% of all email address domains; 0.01% of all commits were authored by these email address domains). This process resulted in a list of 3,170 unique email address domains.

In the issue dataset, we used the 'merged list' to classify users whose affiliations are findable via their profiles into professional affiliation categories. Next, we matched the users who didn't provide their affiliations in profiles but whose email addresses revealed their affiliation with our list of employers' email domains and affiliations. Finally, we classified the users who engaged in issues but did not reveal their affiliation, email address or personal domains as 'non-affiliated.'

*Data analysis: commits and issues*

We conducted a descriptive statistical analysis of commits to compare the contributions of researchers and users in other categories in terms of owning repositories and committing. In addition to commits, we used Source lines of code (SLOC) as a metric.

We compared engagement with issues by researchers and other categories to reveal what specific tasks were performed by researchers. We also compared the roles of participants involved in issue attribution and resolution to understand the roles of researchers, compared to other groups. We investigated issues featuring the term 'bug[s]' or labelled as 'bug[s]' in depth to probe the workloads of different categories of maintainers.

We categorised issue labels and counted how much different user affiliation groups engaged with them. To categorise labels, we counted the frequency of labels and sorted them by descending frequency, then randomly selected 50% of issues featuring each label and reviewed the titles and content of issues closely to understand the meanings of labels. Finally, we merged labels sharing identical or similar meanings together. The qualitative reading was conducted by two researchers separately, who discussed initial findings to develop a cohesive coding strategy. The two researchers then worked together to merge labels into categories, and a third researcher examined the categories and closely read a random selection of issues to ensure the classification was consistent and valid.

Although issues are not explicitly designed to facilitate non-coding-related communications, we observed that researchers do use them to seek new maintainers and to express themselves. To understand the motivations behind maintainer recruitment (and how these transitions are negotiated), as well as the challenges of maintaining, we purposively selected a sub-corpus of issues for thematic analysis. We chose to use thematic analysis because of its flexibility when identifying patterns across text (Lawless & Chen, 2019).

Rather than aiming for statistical representativeness, our selection prioritised repository issues explicitly addressing labour-related concerns. This allowed us to examine the challenges of repository governance and negotiations of responsibilities involved in sustaining R package projects. By filtering issue titles using the keywords 'maintainer,' 'developer,' and 'contributor,' we identified 23 discussions that explicitly documented maintainer recruitment and negotiation processes—often titled with phrases such as 'New Maintainer Wanted :-)' or 'New maintainer of package.'

A deductive coding approach was applied (Braun & Clarke, 2021). After familiarising themselves with the corpus, two researchers manually coded the text separately, grouping codes into potential themes. All three researchers then met to review this round, and a consensus was reached. The two initial researchers recoded separately, grouping codes into potential themes. The team then met to finalise the list of themes.

*Limitations*

We did not investigate all R package development, but focused on packages published on CRAN, whose repositories were available at the time of data collection. Since CRAN is the central distributor of R packages, our study represents a reasonable coverage of R package development. Our selection of site may introduce biases given CRAN packages undergo stricter checks, so they overrepresent well-maintained projects which are more visible, attracting higher engagement, and often originate with institutional developers due to compliance requirements. CRAN packages tend to be older and more stable, which may skew results toward mature projects rather than fast-evolving GitHub-only repositories.

Validity issues when using email address domains as proxies for professional affiliation raise concerns. Not only do some firms require employees to use their professional email address when contributing to projects as part of their paid employment; firms may also prohibit their employees from using their professional email addresses when volunteering.[6] At the same time, GitHub users may forget to comply with these rules when they commit and create or comment on issues to different repositories. Another complicating factor is that some developers may want to keep their identities consistent across platforms (e.g., CRAN and GitHub) and over time; however on GitHub identities are indicated by usernames, whilst on CRAN they are indicated by email addresses. Finally, some developers may not be aware that GitHub permits the use of different email addresses which correspond to different purposes such as employment or voluntary work but are linked to the same usernames. All this may lead to inaccurate determination of users' affiliations. Considering the large scale of this research, it was infeasible to verify affiliations, for example by requesting individuals confirm them, so identifying affiliations via email address domains remained our preferred heuristic. For the same reason, even in cases when we determined that users were working in research institutions - by visiting their personal websites and verifying affiliating entities' websites - we classified them as non-affiliated.

Our dataset comprises a large proportion of users classified as non-affiliated. Some email address domains are no longer searchable, as the lifetime of URLs is approximately 10 years (Kern et al., 2020). We address possible behavioural explanations in our findings. This limitation is often a feature of large-scale studies.

---

[6] See https://docs.github.com/en/account-and-profile/setting-up-and-managing-your-personal-account-on-github/managing-your-personal-account/best-practices-for-leaving-your-company

# Findings

## *Non-affiliated contributors and researchers lead R package development*

Researchers took more active roles in creating repositories and committing. Table 3 shows that (excluding non-affiliated users) researchers are by far the most frequent owners of repositories.

**Table 3. Proportion of repository owners by category**

| Type of affiliation | No. of owners (%) |
|---|---|
| Non-affiliated | 4,611 |
| | 52% |
| Research | 2,723 |
| | 31% |
| Firm | 1,128 |
| | 13% |
| Foundation | 328 |
| | 4% |
| Government | 134 |
| | 2% |
| **Total** | **8980** |
| | **100%** |

Table 4 presents findings of which categories contributed most. Column 1 features an organisational analysis of code contributions (from both commit authors and committers): research organisations are by far in the lead. Non-affiliated users were excluded as we were unable to systematically determine their professional status. The rest of Table 4 lists individuals, so it includes non-affiliated contributors, whose numbers are much larger than other categories. We have no way of comprehensively ascertaining who these contributors are or why they are not using professional email addresses when contributing. One possible explanation is that these are volunteer (unpaid) commits made by firm employees who do not wish their employers to be made aware of their professionally unsanctioned contributions.

Another explanation is that users may be accidentally contributing with their personal email address.

**Table 4. Numbers of institutions, commit authors, commits and source line of code (SLOC) by user category, descending by N. contributors.**

| Type of affiliation | No. of organizations | No. of commit authors | No. of commits authored (%) | No. of commits authored weighted (%) | SLOC (%) | SLOC weighted | Mean SLOC of commits authored | No. of committers (%) | No. of commits committed (%) |
|---|---|---|---|---|---|---|---|---|---|
| Non-affiliated* | NA | 17,173 | 1,606,532 | 1,269,245 | 1,783,875,952 | 1,187,771,862 | 1,110 | 10,781 | 1,468,440 |
|  |  | 65% | 68% | 69% | 64% | 61% |  | 64% | 62% |
| Research | 2,398 | 5,802 | 515,936 | 373,508 | 808,518,602 | 599,401,019 | 1,567 | 3,936 | 451,693 |
|  |  | 22% | 22% | 20% | 29% | **31%** |  | 23% | 19% |
| Firm | 1,427 | 2,738 | 144,565 | 118,503 | 142,317,816 | 119,368,971 | 984 | 1,541 | 347,869 |
|  |  | 10% | 6% | 6% | 5% | 6% |  | 9% | 15% |
| Foundation | 339 | 566 | 58,959 | 48,468 | 35,682,992 | 31,954,429 | 605 | 319 | 62,634 |
|  |  | 2% | 3% | 3% | 1% | 2% |  | 2% | 3% |
| Government | 76 | 258 | 28,684 | 22138 | 31,652,894 | 18,989,858 | 1,104 | 182 | 24,040 |
|  |  | 1% | 1% | 1% | 1% | 1% |  | 1% | 1% |
| **Total** | 4,240 | 26,537 | 2,354,676 | 1,831,862 | 2,802,048,256 | 1,957,486,139 | 1,190 | 16,759 | 2,354,676 |
|  |  | 100% | 100% | 100% | 100% | 100% |  | 100% | 100% |

*Note: Since these users didn't state their affiliations, we couldn't determine their affiliations or the number of affiliations they worked for.

Weighted by the number of stars using quantile binning method to assign weight 0.2, 0.4, 0.6, 0.8, 1.0.

Researchers are much more active in authoring commits than other identifiable categories (22%, vs 6% for firm employees and 3% for foundation employees). In addition, researchers contribute more Source Lines of Code (SLOC) on average than all other categories of contributors (including non-affiliated), showing that their contributions are significant. To account for differences in package value, we weighted coding contributions (commits and SLOC) based on repository popularity, measured by the number of repository stars on GitHub. Using quantile binning, we grouped stars into five categories and assigned fixed weights: 0.2 for 0–1 stars, 0.4 for 1–5, 0.6 for 5–13, 0.8 for 13–41, and 1.0 for 41–23,128. There were no significant changes from unweighted results.

At the individual level, we calculated the median number of commits and SLOC authored by developers in each category (Table 5). These results were confirmed by pairwise comparison.[7] Researchers and government employees were the most active contributors in terms of number of commits. The difference in SLOC between researchers (median = 1,844) and government-employees (median = 2,976) is significant. Comparisons with other groups at the individual level shows that researchers are significantly more active than non-affiliated users, firm employees and foundation employees. The prominent impact of a small number of government employees reflects their role as data providers. Of the 623 commits whose SLOC is above the mean (1,190), three quarters originated from the US's National Oceanic and Atmospheric Administration (35%) and United States Geological Survey (30%). Other global government agencies – e.g., statistics, agriculture, and health – also frequently release (often geo-tagged) data.

---

[7] We compared the commits and SLOC of each category to all other categories (pairwise comparisons) using the Wilcoxon rank sum test with continuity correction. The pairwise results show that although government employees (median = 12) authored slightly more commits than researchers (median = 11) as seen in Table 4, the difference between these groups is insignificant.

**Table 5. Median number of commits and SLOC by user categories**

| Type of affiliation | Median of commits by individuals (sd) | Weighted Median of commits by individuals (sd) | Median of SLOC by individuals (sd) | Weighted Median of SLOC by individuals (sd) |
|---|---:|---:|---:|---:|
| Non-affiliated | 6<br>492.98 | 4.0<br>447.42 | 714<br>2209620.22 | 451<br>738953.7 |
| Research | 11<br>328.60 | 7.0<br>268.60 | 1844<br>2531671.88 | 990<br>2282071.1 |
| Firm | 4<br>223.62 | 3.0<br>196.42 | 313<br>458195.70 | 242<br>428032.7 |
| Foundation | 4<br>763.15 | 3.0<br>648.10 | 391<br>776817.45 | 284<br>746451.8 |
| Government | 12<br>264.19 | 8.9<br>219.58 | 2976<br>841598.31 | 1831<br>380992.7 |

*Engagement with issues and comments*

Researchers were the second most active category in terms of *creating* issues (N:55,628, 19.5%), of having issues *assigned* to them (N:8,393, 22.9%), of *closing* issues (N:47,188, 19.2%), and of the number of *comments* (N:68,769, 18.0%) they made to the issues in our dataset. The largest category was non-affiliated users, who engaged in more than 60% of issue *creation* (N:180,229, 63.2%), issue *assignation* (N:22,177, 60.5%), issue *closure* (N:148,484, 60.5%), and issue *commentary* (N:238,271, 62.4%). There were more government employees participating in issues than foundation employees, but foundation employees' engagement was superior to that of government employees (Table 6).

The per capita engagement of individual researchers was average. While the total participation and number of foundation employees is small, they were most actively engaged in creating issues (20 issues per capita), closing issues (46 per capita) and commenting on issues (71 per capita), though this likely reflects their early participation. Firm employees were, on an individual basis, the second most likely to create issues (9 per capita), the first - similarly to non-affiliated users - to be assigned issues (14 per capita), the first to close issues (23 per capita) and make comments (17 per capita). Firm employees were most likely (together with non-affiliated) to assign issues to others (16 per capita) and excluding Foundation employees (N:302) the most likely to close issues (23 per capita, against 15 per capita for researchers and non-affiliated). The top ten firms in terms of their employees most assigning users to others include data science and analytics companies such as R Square Academy and Mazama as well as RTE, Microsoft and Philips. Finally, Table 7 shows that in addition to being proportionately the highest owners of repositories (5%, against 4% on average for all contributors), 2% of researchers were members (against 1% on average), 5% were collaborators (against 3% on average) and 13% were contributors (against 12% on average).

*Researchers with key responsibilities were more active when dealing with issues*

Researchers who were owners, members, collaborators or contributors engaged in issues more than the average (Table 7). In terms of total numbers of owners, members, collaborators and contributors of repositories, researchers (N:2,469, 24%) contributed the most issues after non-affiliated users (N:6,039, 59%), followed by firm employees (N:1,398, 14%), foundation employees (N:177, 2%) and government employees (N:174, 2%).

The proportion of foundation employees with official roles engaging in issues was 43%, which is higher than the proportion of researchers (26%). However, given the number of researchers with official roles (N:2,469) is much higher than that of foundation employees with official roles (N:177), combining the total number of researchers with official roles and the issue engagement percentage shows that researchers were most active when engaging with issues.

**Table 6. Number of users who created, commented on, assigned, were assigned and closed issues by category**

| Type of affiliation | No. issues created<br>No. of users<br>Issue per capita | No. issues assigned<br>No. of users<br>Issue per capita | No. issues assigned by others<br>No. of users<br>Issue per capita | No. issues closed<br>No. of users<br>Issue per capita | No. comments<br>No. of users<br>Issue per capita |
|---|---|---|---|---|---|
| **Non-affiliated** | 180,229<br>31,548<br>6 | 20,530<br>1,322<br>16 | 22,177<br>1,587<br>14 | 148,484<br>9,982<br>15 | 238,271<br>19,217<br>12 |
| **Research** | 55,628<br>8,038<br>7 | 7,522<br>572<br>13 | 8,393<br>671<br>13 | 47,188<br>3,160<br>15 | 68,769<br>5,448<br>13 |
| **Firm** | 38,668<br>4,511<br>9 | 5,169<br>332<br>16 | 5,047<br>356<br>14 | 38,769<br>1,687<br>23 | 53,610<br>3,231<br>17 |
| **Foundation** | 6,008<br>302<br>20 | 275<br>35<br>8 | 343<br>40<br>9 | 6,651<br>145<br>46 | 16,068<br>225<br>71 |
| **Government** | 4,862<br>549<br>9 | 640<br>50<br>13 | 691<br>58<br>12 | 4,244<br>211<br>20 | 4,990<br>400<br>12 |
| **Total** | 285,395<br>44,948<br>6 | 34,136<br>2,311<br>15 | 36,651<br>2,712<br>14 | 245,336<br>15,185<br>16 | 381,708<br>28,521<br>13 |

**Table 7. Number of owners, members, collaborators, contributors and users with no official role engaging in issues in R packages repositories**

| Type of affiliation | N. owners | N. members | N. collaborators | N. contributors | N. no role | Total |
|---|---|---|---|---|---|---|
| Non-affiliated | 1,185 | 360 | 997 | 3,497 | 28,925 | 34,964 |
|  | 3% | 1% | 3% | 10% | 83% | 100% |
| **Research** | **462** | **232** | **495** | **1,280** | **7,026** | **9,495** |
|  | **5%** | **2%** | **5%** | **13%** | **74%** | **100%** |
| Firm | 191 | 124 | 190 | 893 | 3,912 | 5,310 |
|  | 4% | 2% | 4% | 17% | 74% | 100% |
| Foundation | 23 | 18 | 31 | 105 | 239 | 416 |
|  | 6% | 4% | 7% | 25% | 57% | 100% |
| Government | 29 | 26 | 30 | 89 | 493 | 667 |
|  | 4% | 4% | 4% | 13% | 74% | 100% |
| **Total** | **1,890** | **760** | **1,743** | **5,864** | **40,595** | **50,852** |
|  | **4%** | **1%** | **3%** | **12%** | **80%** | **100%** |

*Engagement with issue labels*

In our issue dataset, 68,305 issues had labels, and 217,090 did not. Besides the nine default labels, contributors engaged with 2,618 custom-made labels, so the total number of unique labels used was 2,627.

R community developers relied on a variety of conventions to name issue labels. For example, 2,998 issues featured labels beginning with 'type' (67 unique labels), and 220 issues featured labels beginning with 'topic' (46 unique labels). These terms are redundant (labels are categorising tools), so neither 'topic' nor 'type' provide useful information for discussion or task tracking. Emojis were also used to create labels: there were 1,927 issues featuring 57 unique labels containing at least one emoji (identified with the 'remoji' package).

Our issue dataset comprised multiple labels sharing functional similarity. For example 'features,' 'features requests' and 'new features' were used as equivalents to 'enhancement,' and 'docs' was used as equivalent to 'documentation.' To better map how different user categories engaged with labels, we aggregated similar labels into *label groups*. For example we aggregated 'bug,' 'bug report,' 'bugs,' 'upstream' and 'bug 🐛' into the 'bug' label group, regardless of whether they were default labels (provided by GitHub) or not. In addition to label groups relating to default labels such as 'bug,' 'documentation' and 'enhancement,' eleven label groups emerged from our dataset: priority, progress, test & review, dependencies, level of difficulty, data related, discussion, release, task, reprex, and pkg related.

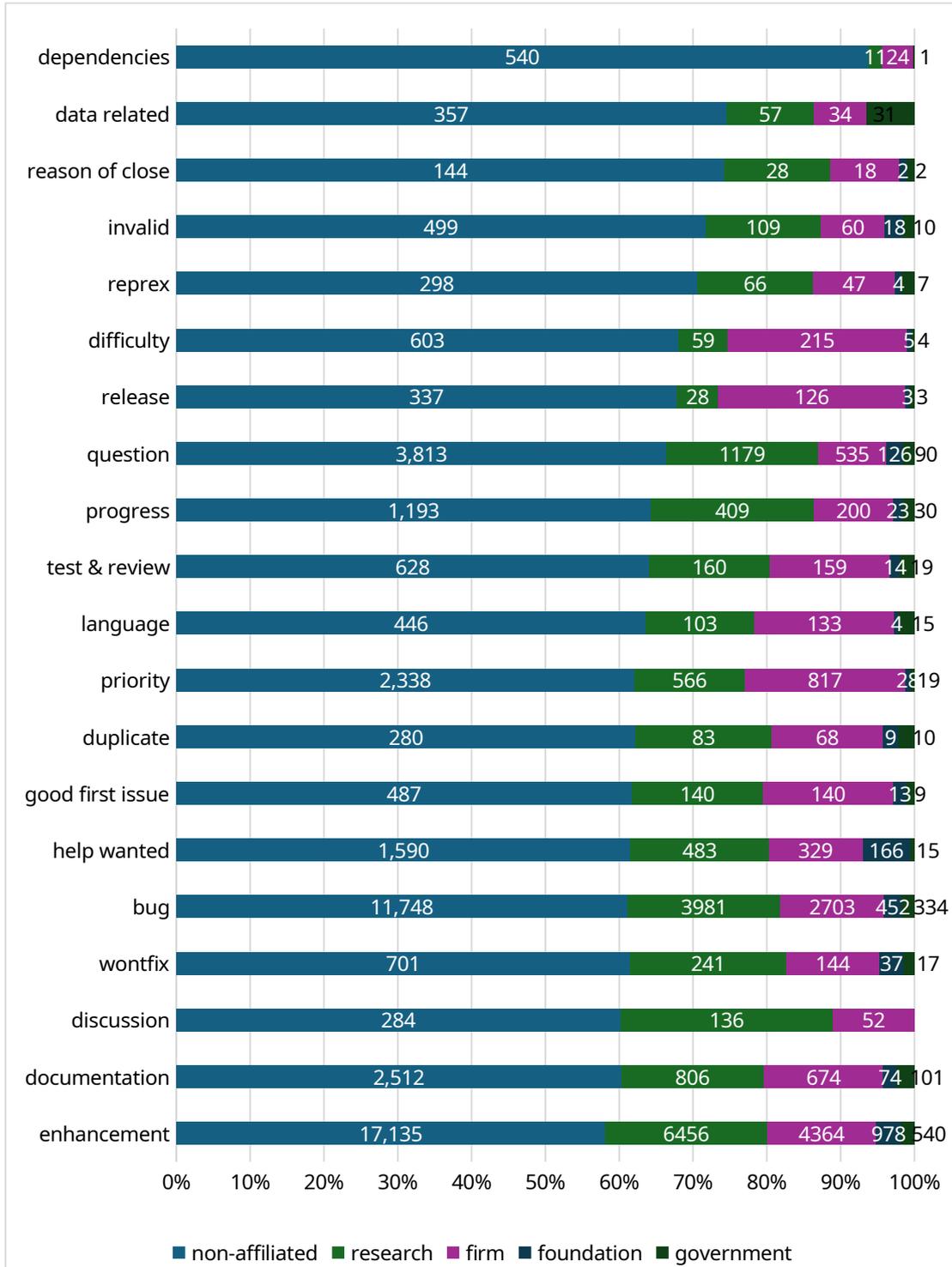

**Figure *1* Number and percentage of issues featuring labels in top 20 most frequently created label groups by user category**

Figure 1 shows the frequency of issue creation featuring top-20 label groups. The most issues appeared in the following label groups: 'enhancement' (N:29,473), 'bug' (N:19,218), 'question' (N:5,743) and 'documentation' (N:4,167). All of these terms were used in proportions roughly equivalent to the size of each category of users. In contrast, other label groups were used disproportionately by specific categories. The label groups most *used* by non-affiliated users, in relation to other categories, were 'dependencies' (94%), 'invalid' (72%), 'data related' (75%), 'reprex' (71%), and 'reason of close' (74%). We understand them as indicating that uppermost concerns were methodological, regarding issue quality. 'Dependencies' indicates the problem is due to the upstream packages the repository is built on. 'Invalid' signifies that the issue is not an actual bug, or is not aligned with the repository's scope, or the problem is caused by user error. In the case of 'data related,' the issue was not a bug, but users encountered problems such as data formatting or dataset availability. The 'reprex' R package generates reproducible examples which can be shared online to seek help or report a bug.[8] The R community encourages its use for clearer issue reporting. Issues labelled with 'reprex' by maintainers initially lacked properly formatted reproducible examples. In such cases, maintainers labelled them 'regrex' and advised issue creators to rewrite their examples using 'reprex' to better diagnose problems. 'Reason of close' indicates whether a closed issue was solved, deemed irrelevant, or postponed for future consideration. The label groups most *used* by researchers in relation to other categories were 'enhancement or feature request' (22%), 'bug' (21%), 'help wanted' (19%), and 'discussion' (29%). We interpret these label groups as referring to deliberative attempts to find the best solutions in a collaborative fashion. Finally, the label groups most *used* by firm employees were 'documentation' (16%), 'priority' (22%), 'difficulty' (24%), and 'release' (25%). Our interpretation is that these label groups concern the timely and practical delivery of results and requests for help via the provision of documentation.

---

[8] See https://reprex.tidyverse.org/

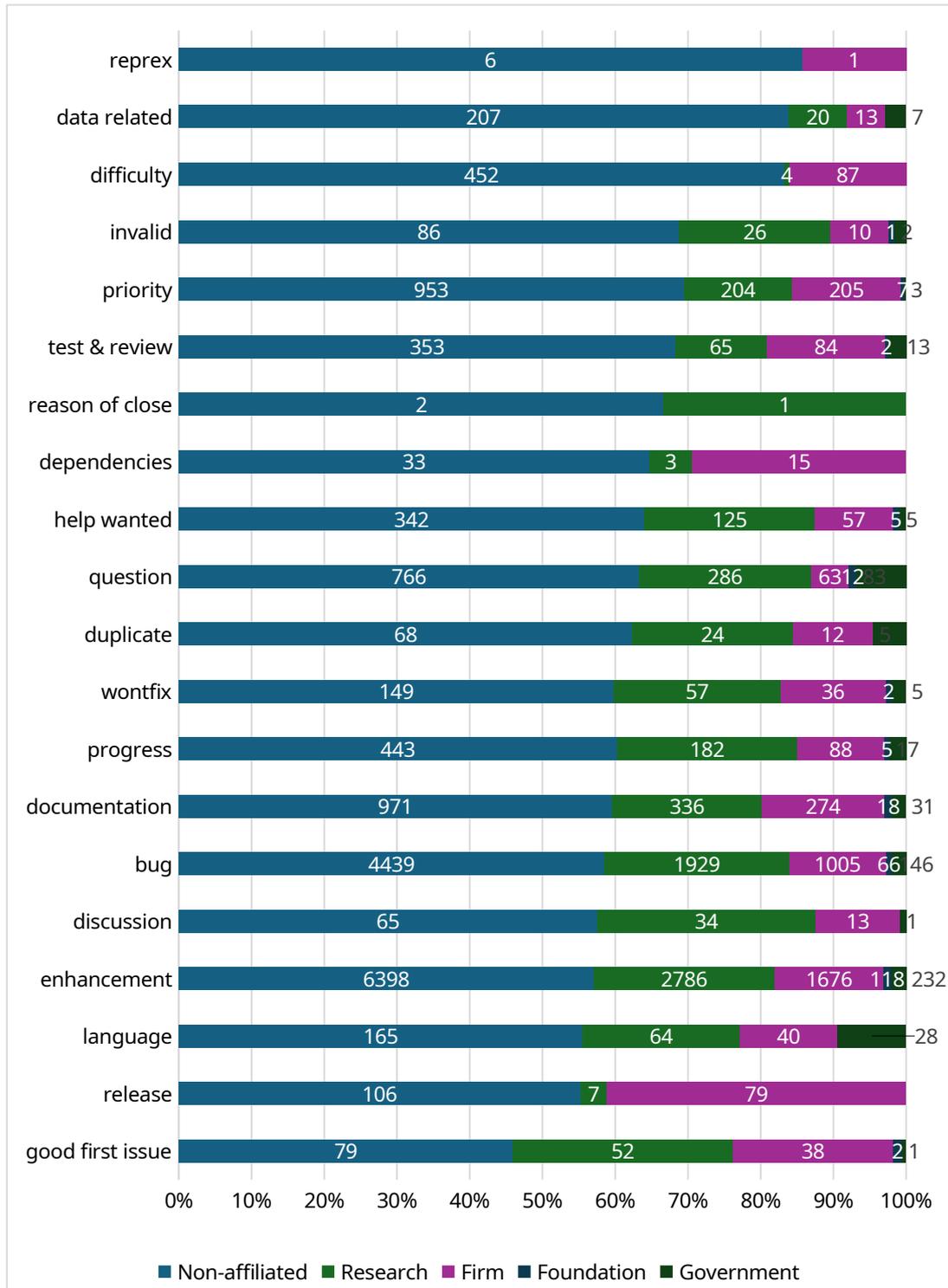

**Figure 2** Number and percentage of times issues are assigned by top-20 label group and user category

Figure 2 shows the label groups comprising issues that were most frequently *assigned* to non-affiliated users in relation to other categories were 'priority' (69%), 'data related' (84%), 'reprex' (86%), and 'invalid' (69%). In our interpretation, this means that, similarly to the issues *used* by non-affiliated contributors (see Table 8), all these issues (apart from 'priority') raised quality concerns. The label groups comprising issues that were most frequently *assigned* to researchers in relation to other categories were 'enhancement' (25%), 'question' (24%), 'progress' (25%), and 'good first issue' (30%). The latter is a classic developer phrase which aims to encourage and welcome new contributors. 'Enhancement' refers to an improvement request whereas 'question' indicates users believe a new development direction is being proposed, in contrast to an issue being raised. Finally label groups most frequently *assigned* to firm employees in relation to other categories were 'documentation' (17%), 'release' (41%), 'good first issue' (25%), and 'dependencies' (29%). We understand these as denoting a professional production environment whereby these users are being asked to improve documentation for releases which only they can authorise.

In sum, parallel worlds emerge from our data: the communal world of unaffiliated users, who are often assigned 'amateur' contributions of questionable quality, thereby maintaining the community's norms and culture; the caring world of researchers, who assist others; and the professional world of firm employees, ruled by tight deadlines.

### *Challenges: poorly labelled issues, burnout and succession management*

Bug reporting is crucial in software development. We found there was a mismatch between issue titles where issue creators had included the term 'bug[s]' and the issues labelled as 'bug[s]' (Table 10). Of 1,431 issues whose titles featured the term 'bug' or 'bugs,' 70% (N:997) were labelled as bugs, and 30% (N:434) were not labelled as 'bug' or 'bugs.' Meanwhile, 24% (17,443) of issues that did not have the terms 'bug' or 'bugs' in their titles were labelled as 'bugs.'

These results show that 'bug' in issue titles were more often than not incorrectly labelled, so developers had to navigate many issues to identify and triage bug reports. In other words, this created unnecessary work and complicated the task of responding to issues for maintainers.

**Table 10. Number of issues with titles (not) featuring bug / error, and (not) labelled as bug report**

|  | Labelled as bug | Not labelled as bug | Total |
|---|---|---|---|
| **Issues with titles featuring bug** | 997 | 434 | 1,431 |
| % in row | 70% | 30% | 100% |
| % in column | 5% | 1% | 2% |
| **Issues with titles not featuring bug** | 17,443 | 54,976 | 72,419 |
| % in row | 24% | 76% | 100% |
| % in column | 95% | 99% | 98% |
| **Total** | **18,440** | **55,410** | **73,850** |
| % in row | 25% | 75% | 100% |
| % in column | 100% | 100% |  |

We detected two other maintainer recruitment challenges through thematic analysis: overwork among existing maintainers and the negotiation of onboarding conditions. The first theme highlights researcher concerns about workload distribution in FLOSS projects. One maintainer expressed this challenge explicitly: 'I maintain too many packages, would be very happy if you are willing to help with maintaining this one!' This need for support directly influences onboarding discussions, where repository access, technical training, and role definitions are negotiated. Newcomers typically start with priority tasks and gradually take on more responsibilities. However, existing maintainers have varying support availabilities. As one maintainer noted, 'I'm available to answer questions, but I won't have time to write any code.' Another key aspect of negotiation is credit allocation. A structured hierarchy typically determines how contributions are recognised. Maintainers follow clear rules, ranging from authorship in package articles to acknowledgments. This is particularly apparent at key transition moments in the development process, such as when existing package maintainers discuss with new or 'replacement' maintainers how recognition, in the form of authorship, will be allocated. We provide an example in the next paragraphs, in which a repository owner emphasised the importance of updating contributor details:

> *Previous maintainer: 'Everything should be transferred over. Make sure you update DESCRIPTION with the new authorship details.'*
>
> *New maintainer: '@ Previous maintainer I'm planning to change the DESCRIPTION file soon. Do you want to remain an author for this package or are you happy to be taken off the project entirely?'*
>
> *Previous maintainer: 'So, the way this should work is that my role in DESCRIPTION should be 'author' for the package… I don't need to be an author on any papers, etc. but obviously an acknowledgement or citation to the package itself would be appreciated.*

This example was extracted from a series of issue comments which occurred in May - June 2018 when a new maintainer was onboarding and the previous maintainer stepped down as a collaborator. To protect user privacy, any identifying details have been removed.

Such negotiations over the attribution of symbolic, rather than economic value raise the issue of 'intellectual monopolisation,' i.e., the perpetuation of a practice in which many co-create value and co-produce knowledge that ends up disproportionately captured by a few giants (Rikap, 2024).

## DISCUSSION

Our results reveal the role played by researchers in creating and supporting critical infrastructure which underlies the platform economy. Researchers are the most frequent owners of our selected R package repositories (RQ1), the second-most active commit authors to our selected R package repositories (RQ2) and they tend to take a more active role than firm employees in addressing defective code (as proxied by responding to issues) in the R software ecosystem; they also more actively deal with issues, are assigned issues more than other categories, and provide more help to users experiencing non-bug related issues than other categories; finally, they have to contend with improperly titled or labelled issues (RQ3). This is all the more striking in comparison to the characteristics of firm employees who were the most active category in terms of creating and closing issues and were mainly concerned with the 'release' of packages. We interpret our analyses of which issues different categories of contributors use, or are being assigned, as signifying that researchers are being given tasks by firm employees.

Such tensions are arguably inherent in academic work, particularly in academic work which extends beyond academia into the world of volunteer labour, and have been present since the start of computer hacking. Academic labour featured prominently in early accounts of computer hacking, with portrayals of unsung 'heroes' in famous university settings such as the MIT Artificial Intelligence Lab (Levy, 1984). Significantly, the foundational event of the FLOSS movement occurred when a researcher quit MIT because of frustration with proprietary software (Stallman, 1985). For this former researcher, Richard Stallman, whose opinions have historically been highly influential in this debate, FLOSS produced by free labour would always be preferable to non-free software produced by decently paid waged labour: the defence of the 'four freedoms' (to use, study, modify, and redistribute the software) mattered more than the fair distribution of profits stemming from software development (Broca, 2018).

This normative stance was made very early in the history of free software. The GNU Manifesto asserts that 'extracting money from users of a program by restricting their use of it is destructive' (Stallman, 1985). Over the years, every time he was asked to comment on the valuation of IT firms and the fact that they benefit from unpaid voluntary labour, R. Stallman gave the same answer: these issues are secondary. They are mainly 'a distraction from what really matters: that these programs (e.g., FLOSS) are available for everyone to use in freedom and community' (Stallman, 2018). In other words, the free software movement should consider software as *resources* upon which users have certain *rights*, not as *products* of a *labour* that deserves *monetary retribution* (Broca, 2018).

While computer engineers who are later employed by private firms can monetise their voluntary labour, we argue that the FLOSS ideology which does not systematically link labour to remuneration effectively legitimises Big Tech firms' appropriation of academic labour. In contrast to the traditional academic model in which researchers create something, are 'paid' in the form of cited publications, and move on, with university rents researchers provide a key service, gratis, for industry users of statistical software and have no guarantee that their contribution will be recognised in the form of acknowledgements. Furthermore, lack of support and challenges such as overwork, incorrect issue reports, recruiting suitable replacements and organising the allocation of symbolic rewards for new maintainers' labour all take a toll on researchers who develop code on GitHub.

The challenges faced by academic FLOSS maintainers point to a wider issue: the lack of comprehensive institutional recognition of FLOSS. After decades of intensive lobbying by Big Tech firms in favour of their hegemonic products and a dominant discourse that, defining innovation solely in terms of private investments and start-ups, delegitimises alternatives in advance, the State's role was mainly to sign license renewals for Big Tech firms. Support for FLOSS was inconsistent. Some objections were eventually raised: if FLOSS are now key digital infrastructure, should security flaws that affect some projects, whose consequences can be harmful, not be of concern?

In the wake of critical vulnerabilities within a Java logging library, Apache Log4j, which were discovered in December 2021,[9] the US Presidency decided to gather stakeholders from US government agencies, Big Tech companies, and FLOSS foundations to improve FLOSS security. Participants discussed 'how to prioritise the most important open source projects and put in place sustainable mechanisms to maintain them,' but these mechanisms did not include a coordinated government response.[10] In contrast, Log4j incited Germany to set up a Sovereign Tech Agency in 2022, which precisely intends to address market failures by supporting FLOSS developers and projects, notably through its Bug Resilience Program.[11]

More radical 'societal' alternatives raise the questions of the recognition of voluntary contributions and the articulation between the cooperative, state, and private sectors in a context of increasing automation and unemployment. In France *Les Économistes Atterrés* and Bernard Stiegler have, for example, proposed variants of 'common labour rights' that would allow those who contribute to the commons to accumulate rights of access to social services (Maurel, 2019). Similarly, the Digital Commons Policy Council, an international think tank for the commons based in Australia released in 2021 a report in which, during a discussion of recognition mechanisms for volunteer labour such as the production of FLOSS, arguments for and against Universal Basic Incomes were explored (O'Neil et al., 2021).

---

[9] RNZ, Apache Log4j: Software flaw 'being actively exploited', CERT NZ warns, 13 December 2021. https://www.rnz.co.nz/news/national/457779/apache-log4j-software-flaw-being-actively-exploited-cert-nz-warns

[10] See https://www.whitehouse.gov/briefing-room/statements-releases/2022/01/13/readout-of-white-house-meeting-on-software-security/

[11] See https://www.sovereign.tech/

Such solutions would imply that developers working in academia and industry join forces and take a stand. The key position occupied in the economy by FLOSS developers theoretically gives them the capacity to influence its direction, through industrial action for example. But can the free software community constitute itself as a political entity that reflects, beyond software, on society as a whole? Can it confront libertarian dogmas, productivist orthodoxies, the obligation to infinitely develop computing power? The meaning and value of work is central to everyone – supporting FLOSS and other volunteer labour, not just that of researchers, will help to better recognise work that is socially useful.

## Acknowledgements

Some of our aims and findings were presented at the following events: Australian Social Network Association Conference, University of New South Wales, Sydney; Political Economy Research Section, International Association for Media and Communication Research, Nanyang Technological University, Singapore; Journées du CIS, Centre Internet et Société (CNRS). We are grateful to the participants for their input. We also wish to thank the anonymous colleagues who reviewed this article for their valuable contributions to improving our argument.

## Funding information

This research was supported by operational funding from the Ford Foundation's Technology and Society program.

# Appendix

## *R Consortium members*

| Tie | Member | Description |
| --- | --- | --- |
| **Foundation member** | R Foundation | Not-for-profit |
| **Platinum Members** | Biogen | Biotechnology company |
| | Genentech | Biotechnology company |
| | Microsoft | Technology company |
| | Posit | Technology company (RStudio) |
| **Silver Members** | American Statistical Association | Professional organization |
| | ASCENT | Data analytics consultancy |
| | Esri | Technology company (GIS) |
| | Google | Technology company |
| | GSK | Biopharma company |
| | Johnson & Johnson | Pharmaceutical company |
| | Lander Analytics | Data analytics consultancy |
| | MERCK | Pharmaceutical company |
| | NOVARTIS | Pharmaceutical company |
| | Novo Nordisk | Pharmaceutical company |
| | Oracle | Technology company |
| | Pfizer | Pharmaceutical company |
| | ProCogia | Data analytics consultancy |
| | sanofi | Pharmaceutical company |
| | Swiss Re | Insurance agency |

Note: data derived from https://r-consortium.org/members